\documentclass[aps,fleqn,twocolumn,prl,reprint]{revtex4-1}
\usepackage{graphicx}
\usepackage{amsmath}

\def\state#1{\big|#1\big>}

\def\overlap#1#2{\big<#1\big|#2\big>}

\def\x{\cdot}
\def\z{0}
\def\om{\omega_{\beta}(t)}
\def\om{\omega}
\def\tight#1{\hspace{-0.3em}#1\hspace{-0.3em}}
\def\tight#1{#1}
\def\matrix#1{\left(\!\begin{array}{cc|ccc}#1\end{array}\!\right)}

\def\e#1{{\rm e}^{#1}}
\def\i{{\rm i}}

\def\I{\textsl{I}}
\def\II{\textsl{II}}
\def\III{\textsl{III}}

\begin{document}
\title{Adiabatic passage to the continuum}
\author{Ulf Saalmann}
\author{Sajal Kumar Giri}
\author{Jan M. Rost}
\affiliation{Max-Planck-Institut f{\"u}r Physik komplexer Systeme,
 N{\"o}thnitzer Str.\ 38, 01187 Dresden, Germany}
\date{\today}

\begin{abstract}\noindent
We demonstrate that by changing the direction of the chirp in VUV pulses one can switch between excitation and ionization with very high contrast, if the carrier frequency of the light is resonant with two bound states.
This is a surprising consequence of rapid adiabatic passage if extended to include transitions to the continuum. The chirp phase-locks the linear combination of the two resonantly coupled bound states whose ionization amplitudes interfere constructively or destructively depending on the chirp direction under suitable conditions.
We derive the phenomenon in a minimal model and verify the effect with calculations for helium as a realistic example. 
\end{abstract}

\maketitle

\noindent
Rapid adiabatic passage is an extremely simple, albeit rather efficient and robust population-transfer technique. 
It has been known for a long time \cite{bl46,makr01} and is explained in review papers \cite{viha+01,go03,me17} and textbooks \cite{sh11}. 
Initially investigated for nuclear magnetic resonance \cite{bl46}, it was experimentally implemented with lasers for the first time in the 70s (a list of experiments can be found in a review \cite{viha+01}).
The necessary time-varying resonance condition was achieved by very different means: Stark shifts in molecules \cite{lo74}, chirping the laser frequency \cite{hama+75} position-dependent Doppler shift of the laser beam \cite{krse+85}, or manipulating coupled waveguides \cite{ouvi+17}.
The selected population of adiabatic states with chirped pulses has been measured ``in situ'' by weak-field ionization of the Stark-shifted states \cite{wopr+06,krba+09}.

In the following we will demonstrate that in the context of intense VUV pulses,
a new possibility opens: namely to use the chirp direction as an effective control for
turning ionization ``on'' or ``off'' although ionization is typically inevitable when atoms are exposed to VUV light.
This is quite surprising since the direction of the chirp does not play any role in traditional rapid adiabatic passage \cite{viha+01,go03,me17,sh11}.

\begin{figure*}
\includegraphics[width=0.9\textwidth]{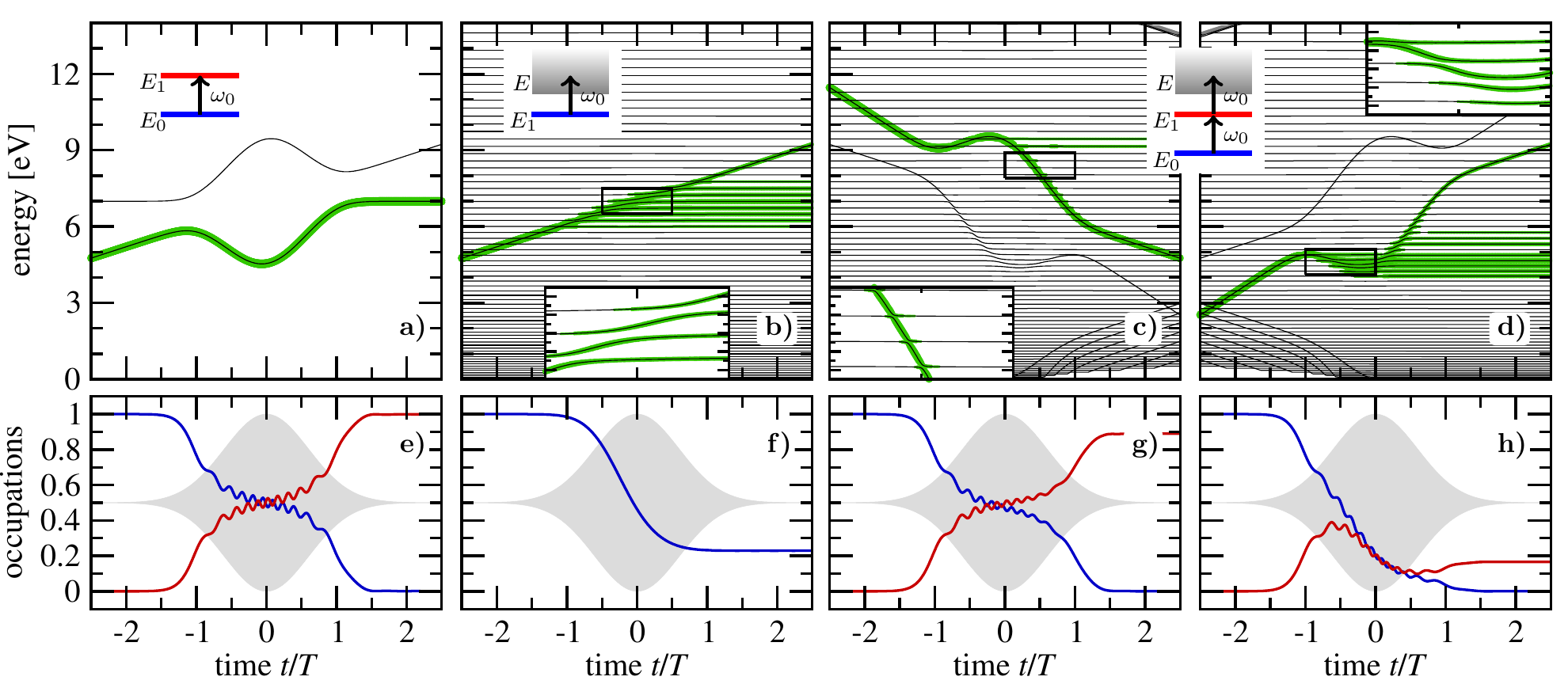}
\caption{Dressed-state description for cases \I, \II\ and \III\ as discussed in the text and sketched in the insets.
Panels {\bf a,\,e}: two-level system with positively chirped pulse (case \I).
Panels {\bf b,\,f}: single level coupled to a continuum with positively chirped pulse (case \II).
Panels {\bf c,\,g} and {\bf d,\,h} two levels coupled to a continuum with with negative and positive chirp, respectively (case \III).
Upper row: Time-dependent energy levels (black lines) and corresponding occupations (green lines with the thickness corresponding to the occupation probability). The dense set of lines in {\bf b}\,--\,{\bf d} represents the discretized continuum.
Lower row: Occupation probability of the two lowest (in {\bf f} only the lowest) field-free states. The gray area shows the envelope of the driving laser pulse.
Time $t$ is measured in units of the chirp-free pulse length $T$.}
\label{fig:model}
\end{figure*}%

We will explain the main mechanism with a minimal model and show subsequently that our findings equally apply to real atoms. 
To this end we consider first an electron in one dimension subject to a soft-core Coulomb potential $V(x)\,{=}\,{-}1/\sqrt{x^{2}{+}1/2}$ and a Gaussian light pulse with the time-dependent vector potential
\begin{subequations}\label{eq:pulse}\begin{align}
{\cal A}_{\beta}(t) & ={\cal A}_{\beta}\,g_{\beta}(t)\cos\big(\phi_{\beta}(t)\big),
\\
g_{\beta}(t) & =\exp\big({-}2\ln\!2\,t^{2}/T_{\beta}{\!}^{2}\big),
\intertext{whose frequency drifts linearly in time}
\omega_{\beta}(t)
& = \frac{{\rm d}}{{\rm d}t}\phi_{\beta}(t)
= \omega_{0}+\frac{4\ln2}{\beta{+}1/\beta}\,\frac{t}{T^{2}},
\label{eq:tdfreq}
\end{align}\end{subequations}
controlled by a dimensionless chirp parameter $\beta$.
The Fourier-limited pulse ($\beta\,{=}\,0$) is characterized by carrier frequency $\omega_{0}$ and length $T$. 
Note, that the strongest chirp is achieved with $\beta\,{=}\,{\pm}1$ and that any chirp stretches the pulse in time to $T_{\beta} = [1{+}\beta^{2}]^{1/2}\,T$ which implies a reduced peak amplitude of ${\cal A}_{\beta} = [1{+}\beta^{2}]^{-1/4}{\cal A}_{\rm max}$, leaving the pulse energy unchanged \cite{pulse}.
We solve the time-dependent Schr\"odinger equation including the light-matter coupling term ${\cal A}_{\beta}(t)\,\hat p$ in the basis of field-free eigenstates \cite[Sect.\,2]{suppl}.
The two lowest states have energies of $E_{0}{=}{-}24.2$\,eV and $E_{1}{=}{-}8.6$\,eV, respectively, corresponding to a transition energy of $\Delta=E_{1}{-}E_{0}=15.6$\,eV.
Although not essential, $E_{0}$ is close to the binding energy of helium.

The two-level system restricted to the two lowest states of the model, constitutes our case \I. It shows upon driving with a chirped laser almost perfect rapid adiabatic passage, cf.\ Fig.\,\ref{fig:model}e.
With a properly chosen carrier frequency, e.\,g.\ $\omega_{0}=\Delta$ as in Fig.\,\ref{fig:model}, the uncoupled dressed states with energies $E_{0}{+}\omega_{\beta}(t)$ and $E_{1}$ would cross. The laser coupling, however, pushes them apart, as shown in Fig.\,\ref{fig:model}a, and thereby suppresses (non-adiabatic) transitions.
As a consequence only one adiabatic state is occupied for all times. Yet, this enables a transition since the adiabatic state changes its character \cite{makr01,viha+01,go03,me17,sh11}.
The sign of the chirp $\beta$ does not play any role.
Indeed, the final occupation probability is identical for pulses ``flipped'' in time \cite[Sect.\,4]{suppl}. 
\begin{figure}[b]
\centerline{\includegraphics[width=0.4\textwidth]{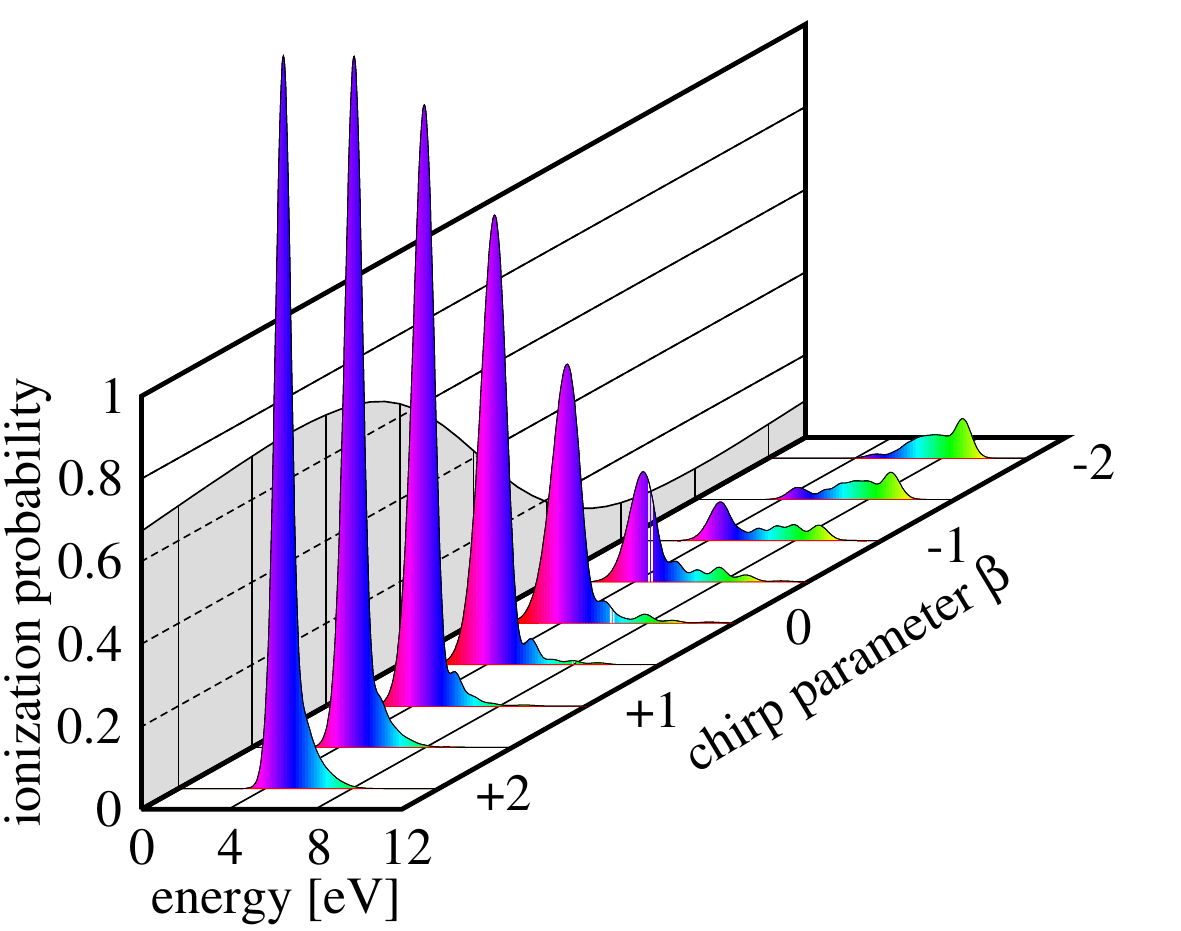}}
\caption{Electron-energy spectra of the model system for various chirp parameters $\beta$.
The pulse parameters \cite{pulse} are $I=3{\times}10^{15}$W/cm$^{2}$, $T=2$\,fs and $\omega_{0}=15.6$\,eV.
Without Stark shift one would expect a peak at $E\,{=}\,7$\,eV.
For negative $\beta$ peaks shift towards higher energies and ionization is increasingly suppressed, most clearly seen from the integrated ionization probability shown as gray-shaded area on the left side.}
\label{fig:spec}
\end{figure}%

What happens if one bound state is replaced by a continuum allowing photo-ionization? We use the same underlying 1D system and laser parameters as before, but remove the ground state and start from the first excited state, constituting case \II.
At a first glance, the dressed state with energy $E_{1}{+}\omega_{\beta}(t)$ seems to intersect the levels of the discretized continuum, see Fig.\,\ref{fig:model}b. However, the diagonalized levels shown are adiabatic levels
and exhibit narrow avoided crossings, facilitating non-adiabatic transitions, see the inset of Fig.\,\ref{fig:model}b. 
They form the most striking difference to the bound two-level system and allow to preserve the initial state, if the 
level crossings are transversed diabatically. Nevertheless, the many possibilities to stay adiabatically in a continuum
level lead to a significant depletion of the initial state. This is seen in Fig.\,\ref{fig:model}f, rendering the bound-continuum ``two-level system'' very similar to the bound-bound one, which is also corroborated by the fact that
reversing the sign of the chirp does not change the depletion. 
Note however, that with a reversed chirp, modified electron spectra may result in the saturation regime, since the matrix elements between initial and continuum states are not identical (but rather often decrease with the corresponding continuum energy). 

In the next and final case \III, we keep the continuum but again add the bound state, we had dropped before.
Through the time-dependent frequency \eqref{eq:tdfreq} and the coupling, two adiabatic states are formed which contain roughly the same amount of the two bound states when the pulse is at full strength around $t=0$,
see Figs.\,\ref{fig:model}g,\,h. 
Yet, for this two-level system which is in contrast to case \I\ additionally coupled to the continuum, reversing the chirp has dramatic consequences: While the negative chirp (Fig.\,\ref{fig:model}c)
leads to a similar effect as in the pure two-level system, namely a nearly complete exchange of the two bound states
apart from small losses to the continuum 
(Fig.\,\ref{fig:model}c), the system almost fully ionizes under positive chirp (Figs.\,\ref{fig:model}d,h). Note that for both chirp directions, the initial state gets fully depleted.

One can get from the insets of Figs.\,\ref{fig:model}c and \ref{fig:model}d a qualitative reason for this drastic difference:
Apparently the upper adiabatic state (referred to as ``$\uparrow$'' in the following)\,--- initially populated for the negative chirp\,---\,hardly couples to the continuum as apparent from the very narrow avoided crossings the inset of Fig.\,\ref{fig:model}c reveals. Comparatively broader avoided crossings (inset of Fig.\,\ref{fig:model}d) indicate that
the lower adiabatic state ``$\downarrow$'', initially populated for the positive chirp exhibits a significant interaction with the continuum, similar to case \II\ with one bound and one continuum state. 
As a consequence the ionization probability differs strongly when reversing the sign of the chirp parameter $\beta$.
For negative $\beta$ one sees hardly any ionization, for positive $\beta$ there is almost 100\,\% ionization. 
We stress that the absolute value of the two pulses in the frequency domain is identical, they differ only by their phase. Hence, the difference in ionization cannot be attributed to a resonance effect: In fact, both pulses are resonant in the same manner. Also, for both chirp directions, the initial state gets fully depleted. One may say that the sign of the chirp decides if the three-level system behaves like the bound-bound (case \I) or the bound-continuum (case \II) system, although for each isolated two-level system the chirp direction does not matter.

Figure\,\ref{fig:spec} summarizes the strong dependence of case \III\ on the chirp, confirming that significant ionization occurs for $\beta\,{>}\,0$ from the lower state ``$\downarrow$'' since the peak energy is $E_{\rm peak}\,{\approx}\,4$\,eV. On the other hand for $\beta\,{<}\,0$, ionization originates from the upper state ``$\uparrow$'' with $E_{\rm peak}\,{>}\,8$\,eV but is strongly suppressed.
Consequently, the total ionization yield (grey-shaded area) suddenly drops as a function of decreasing $\beta$ around $\beta\,{=}\,0$. The two peak positions result from an Autler-Townes splitting \cite{auto55}, from which we
estimate at the maximal field strength ${\cal A}_{\beta}(t{=}0)\,{=}\,{\cal A}_{\beta}$ 
\begin{equation}\label{eq:split}
E_{\downarrow\!\uparrow}=E_{0}+2\hbar\omega_{0}\mp\Omega_{01},
\end{equation}
with the laser frequency $\omega_{\beta}(t{=}0)=\omega_{0}=\Delta$, cf.\ Eq.\,\eqref{eq:pulse}, and the Rabi frequency $\Omega_{01}\equiv\frac{1}{2}|{\cal A}_{\beta}V_{01}|$, where $V_{jj'}$ denotes the coupling matrix element.
We obtain $E_{\downarrow}\,{=}\,3.6$\,eV and $E_{\uparrow}\,{=}\,10.4$\,eV for the parameters of Fig.\,\ref{fig:spec} in good agreement with the peak positions seen there.

Why does a typical two-level adiabatic passage system become extremely sensitive to the chirp direction when coupled to a continuum? An essential dressed-state representation, where the Hamilton matrix is augmented with states ``dressed'' by an appropriate number of photons provides the answer. Summarizing the detailed derivation in the supplement \cite[Sect.\,3]{suppl},
 the minimal states required are the two bound states $\varphi_{0}$ and $\varphi_{1}$, resonantly coupled through $V_{01}$ with $E_{1}-E_{0}=\omega_{0}$ and two states $\varphi_{\rm u}$ and $\varphi_{\rm g}$ in the continuum with the same energy $E$ but different symmetry.
The latter couple among each other through $V_{\rm ug}$ and each of them couples to one bound state via 
 the matrix elements $V_{0\rm u}$ and $V_{1\rm g}$, respectively.
In the dressed-state representation states couple only if they have opposite symmetry (gerade vs.\ ungerade) and if they differ by exactly one dressing photon. Therefore, the effective Hamiltonian matrix constructed from the four states contains only 5 dressed states and can be cast into a form where the 2$\times$2 bound and the 3$\times$3 continuum blocks are diagonalized \cite{suppl}:
\label{eq:matrix}
$$\arraycolsep=2pt
\matrix{%
\tight{E_{0}{+}2\om} & \Omega_{01} & \Omega_{0\rm g} & \z & \z \\[1ex]
\Omega_{10} & E_{1}{+}\om & \z & \tight{\Omega_{1\rm u}} & \z \\[1ex]
\hline \rule{0ex}{3ex}
\Omega_{{\rm g}0} & \z & E{+}\om & \tight{\Omega_{\rm ug}} & \z \\[1ex]
\z & \Omega_{{\rm u}1} & \Omega_{\rm gu} & E & \Omega_{\rm gu} \\[1ex]
\z & \z & \z & \tight{\Omega_{\rm ug}} & E{-}\om \\[1ex]
}
\to
\matrix{%
E_{\downarrow} & \z & \x & C_{\downarrow} & \x \\[1ex]
\z & E_{\uparrow} & \x & C_{\uparrow} & \x \\[1ex]
\hline \rule{0ex}{3ex}
\x & \x & \x & \z & \x \\[1ex]
 C_{\downarrow} & C_{\uparrow} & \z & E & \z \\[1ex]
\x & \x & \x & \z & \x \\[1ex]
}
$$
Here, $\omega$ denotes the instantaneous frequency $\omega_{\beta}(t)$ and the Rabi frequencies are defined as
$\Omega_{jj'}\equiv\frac{1}{2}{\cal A}_{\beta}g_{\beta}(t)V_{jj'}$ with $g_{\beta}$ the Gaussian from Eq.\,(\ref{eq:pulse}b).
We would like to stress that the chirp $\beta$ merely selects the adiabatic state, ``$\downarrow$'' vs.\ ``$\uparrow$'', but is not essential for the subsequent discussion. 
\begin{figure}[t]
\includegraphics[width=0.48\textwidth]{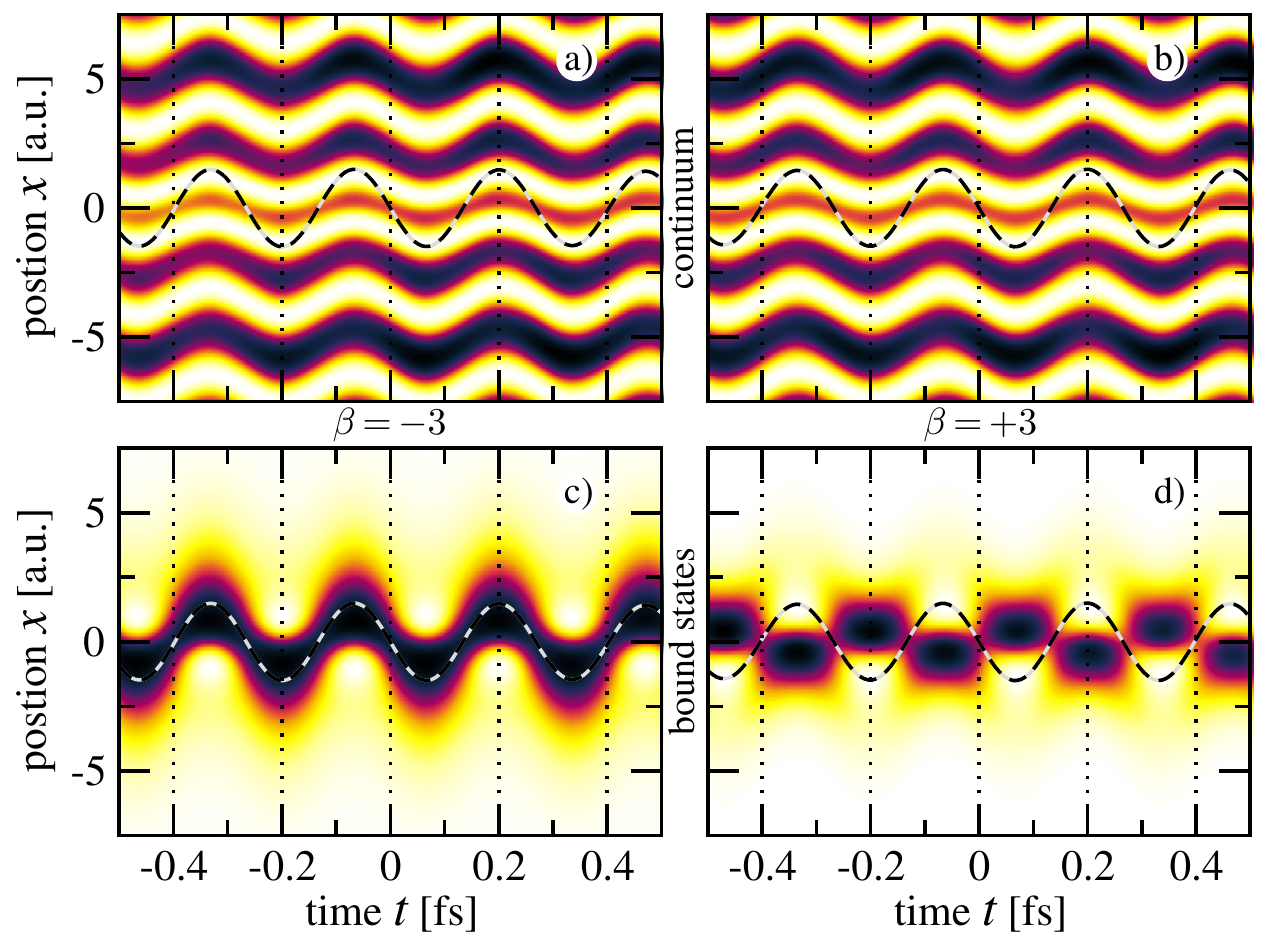}
\caption{Oscillating bound (lower row) and continuum (upper) state density for negative (left column) and positive (right) chirp, respectively. The densities are obtained by solving the TDSE for the 1D model system ($I{=}3{\times}10^{15}$W/cm$^{2}$, $T{=}1$\,fs, $\omega_{0}{=}15.6$\,eV \cite{pulse}) with the initial state being the ground state (lower row) or the gerade state at $E\,{\approx}\,E_{0}{+}2\omega_{0}$ (upper).
The dashed lines indicate the laser's electric field in arbitrary units. (Note that the difference in the chirp is not visible.)}
\label{fig:dens}
\end{figure}%
The coupling terms $C_{\downarrow\!\uparrow}$ read for the case $E_{0}+\omega=E_{1}$ 
\begin{equation}
C_{\downarrow\!\uparrow}=
\frac{\mp\Omega_{0\rm u}\Omega_{\rm ug}-\omega\Omega_{1\rm g}}{\sqrt{2\omega^{2}+4\Omega_{\rm ug}^{2}}}.
\end{equation}
The coupling $C_{\uparrow}$ vanishes for
\begin{equation}\label{eq:supp}
\frac{\Omega_{0\rm u}}{\Omega_{1\rm g}}=
\frac{\omega}{\Omega_{\rm ug}}
\quad\mbox{or}\quad
\frac{{\cal A}_{\beta}}{2}\frac{V_{0\rm u}}{V_{1\rm g}}=
\frac{\omega_{0}}{V_{\rm ug}}.
\end{equation}%
Together with the adiabatic locking to either state ``$\downarrow$'' or ``$\uparrow$'' this explains the very different electron dynamics for positive or negative chirps $\beta$, respectively. That the chirp with direction $\mbox{sign}(\beta)=\pm 1$ creates a locked linear combination in time is illustrated 
in Figs.\,\ref{fig:dens}c and \ref{fig:dens}d with the time-dependent bound-state density in the model which oscillates with the energy difference $\omega_{0} = E_{1}-E_{0}$ of the bound states but is otherwise stationary over the main part of the pulse.

Having established that the chirp generates a locked 
linear combination of the two bound states even in the presence of the continuum (and possibly other bound states), it is tempting to describe adiabatic passage to the continuum as a standard strong-field ionization process \cite{quma+05}
\begin{subequations} \label{eq:ampl}
\begin{align}
\label{eq:psit}
\big|\psi(t)\big\rangle & =-{\rm i}\int^{t}\!{\rm d}t' \hat{U}(t,t') {\cal A}_{\beta}(t')\hat{p}
\big|\psi_{\rm i}(t')\big\rangle,
\intertext{with the initial ``chirp-locked'' state}
\label{eq:amplb}
\state{\psi_{\rm i}(t')} & =\mbox{sign}(\beta)\state{0}\e{\i E_{0}(t')}+\state{1}\e{\i E_{1}(t')},
\intertext{and the (Volkov) propagator}
\label{eq:volkov}
\hat{U}(t,t') & = \e{\i\int^{t}_{t'}\!{\rm d}t''[\hat p^{2}/2+ \hat p\,A_{\beta}\cos(\omega_{0}t'')]}.
\end{align}\end{subequations}
For the latter we fix the vector potential at maximal field strength $A_{\beta}$ and neglect the chirp in the frequency since it changes the photon energy maximally by $|\delta\omega/\omega_{0}|< 0.03$ for our parameters. Indeed, although the chirp differs in the left and right panel of Fig.~\ref{fig:dens}, no differences in the continuum densities are visible. 
\begin{figure}[tb]
\includegraphics[width=0.48\textwidth]{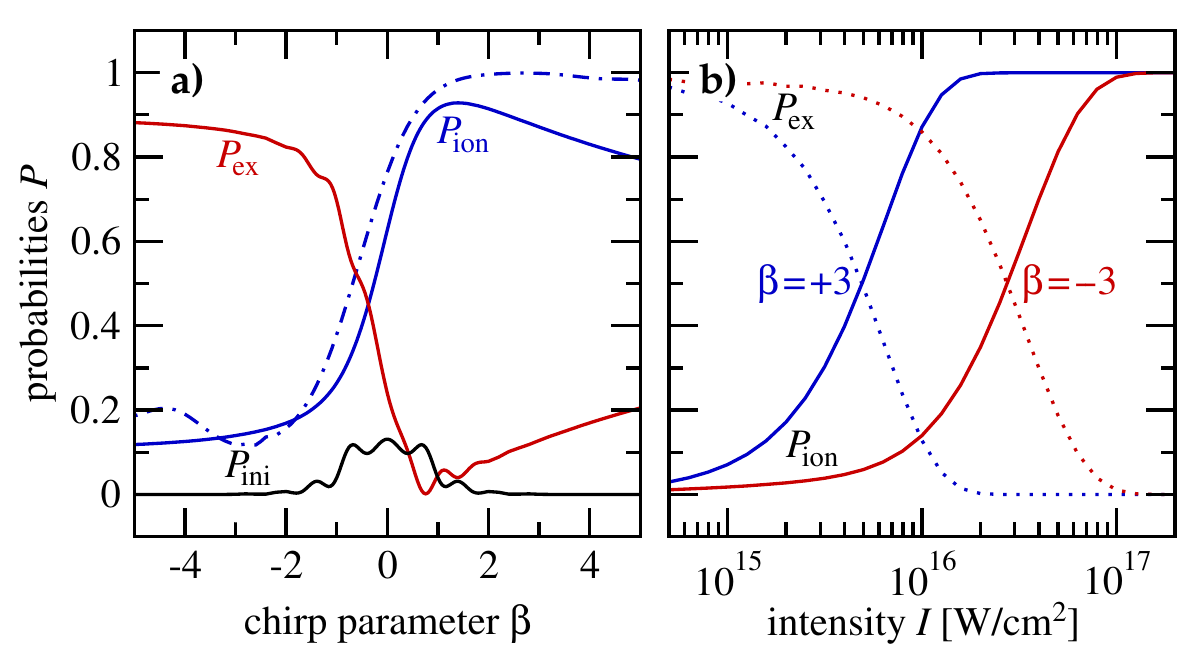}
\caption{Ionization ($P_{\rm ion}$) and excitation ($P_{\rm ex}$) probabilities for Helium as a function of the chirp $\beta$ (panel a, which shows additionally the probability $P_{\rm ini}$ for remaining in the initial state) and as a function of the laser intensity $I$ (b).
The pulse \cite{pulse} had an intensity of $I=10^{16}$W/cm$^{2}$ (a) and a length of $T\,{=}\,3$\,fs (a,\,b).
Additionally we show with a dash-dotted line (a) $P_{\rm ion}$ for the 1D model system discussed in the text.
}
\label{fig:he-probs}
\end{figure}%

With these approximations, we can use Eq.\,\eqref{eq:psit} in order to calculate the amplitude $\overlap{k}{\psi(t)}$ for a specific continuum state with momentum $k$ analytically \cite{suppl}. At resonance $k^{2}/2\,{=}\,E_{0}\,{+}\,2\omega_{0}\,{=}\,E_{1}\,{+}\,\omega_{0}$ and in leading order of the electron-photon coupling parameter
 \begin{equation}\label{eq:couplpara}
\lambda= {A_{\beta}k}/{\omega_{0}},
\end{equation}
the ionization probability $P_\mathrm{ion}^{\rm res} \equiv |\overlap{k}{\psi(t{\to}\infty)}|^{2}$ reads
\begin{equation}\label{eq:ampJ}
P_\mathrm{ion}^{\rm res} \propto \left|\mbox{sign}(\beta)\,\overlap{k}{0}{\lambda^{2}}/{2}
+\overlap{k}{1}\lambda\right|^{2}.
\end{equation}
Obviously, ionization is suppressed for a negative chirp $\beta\,{<}\,0$ if $\overlap{k}{0}{\lambda}/{2}=\overlap{k}{1}$.
It can be shown \cite{suppl} that this condition is equivalent to condition in Eq.\,\eqref{eq:supp} above.
Moreover, $P_\mathrm{ion}^\mathrm{res}$ clearly shows that the suppression or enhancement of ionization in the dressed continuum, dependent on the chirp direction, is due to destructive or constructive interference of the lower bound state $|0\rangle$ having absorbed
two photons $(\propto \lambda^{2})$ and the higher bound state $|1\rangle$ after absorption of one photon $(\propto \lambda)$. 
\begin{figure}[bt]
\centerline{\includegraphics[width=0.48\textwidth]{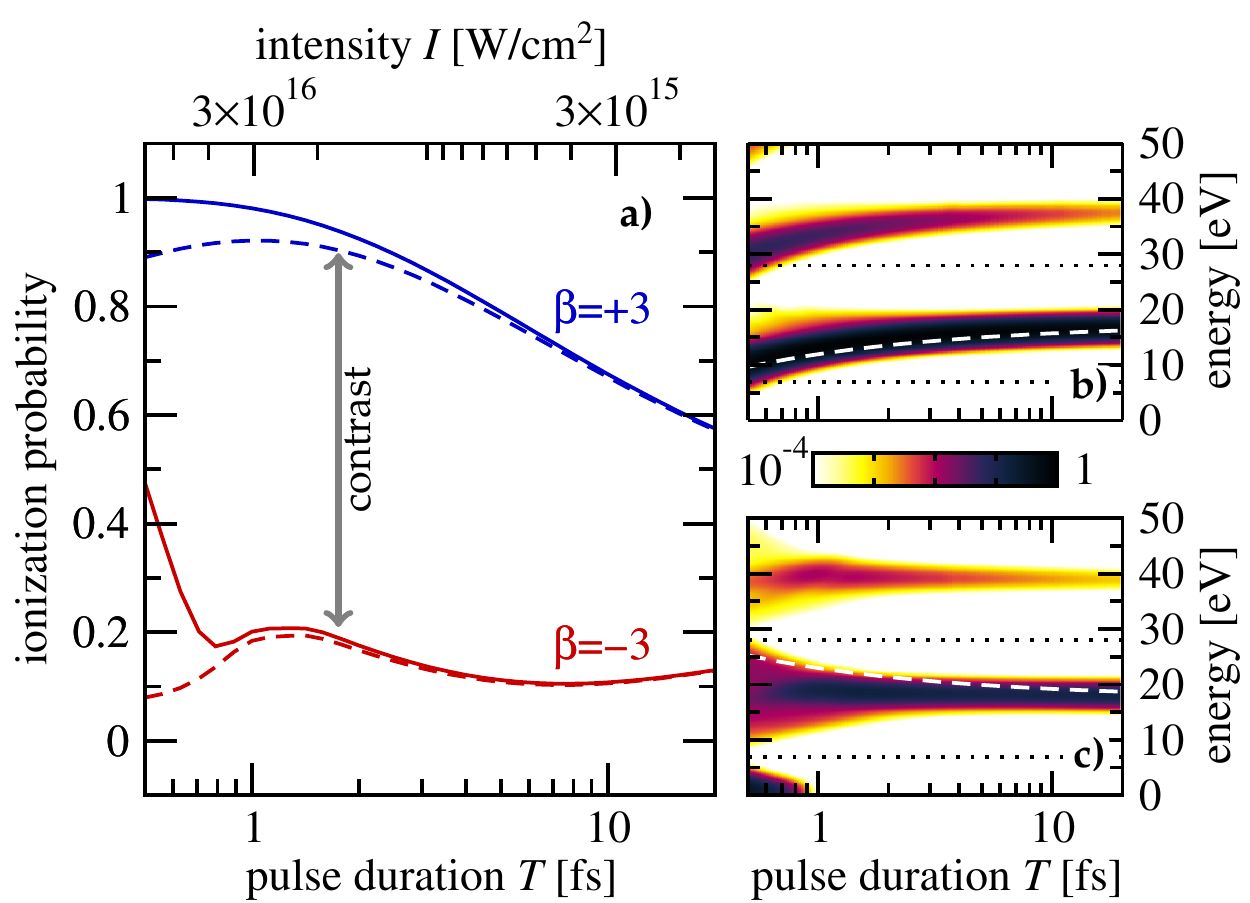}}
\caption{Ionization probabilities as a function of pulse length $T$ for two chirp parameters $\beta\,{=}\,{\pm}3$ (panel a). The pulse energy is kept constant such that $I{\times}T=3{\times}10^{16}$W\,fs/cm$^{2}$.
The dashed lines show the ionization probability with final electron energy $E$ in the ``two-photon region'' $E_{0}{+}3\omega/2\,{\le}\,E\,{\le}\,E_{0}{+}5\omega/2$. This energy interval is marked by dotted lines in the pulse-dependent electron spectra (panels b,c).
The white dashed lines mark the expected spectral peaks from the Autler-Townes splitting of the two $\uparrow\downarrow$ states as given in Eq.\,\eqref{eq:split}.}
\label{fig:he-prob-spec}
\end{figure}%

Having worked out adiabatic passage to the continuum with a minimal model we will finally demonstrate it for realistic 3D systems, whereby $\vec{{\cal A}}_{\beta}(t)\,{=}\,{\cal A}_{\beta}(t)\,\vec{e}_{z}$. To this end we will present calculations for a helium atom within the single-active-electron approximation, certainly applicable for the parameters used. With technical details summarized in the supplemental material \cite{suppl}, we show results in Figs.\,\ref{fig:he-probs} and \ref{fig:he-prob-spec}.
As before, all calculations are done for the resonant frequency $\omega_{0}\,{=}\,\Delta\,{=}\,E_{\rm 2p}{-}\,E_{\rm 1s}\,{\approx}\,21$\,eV, but apply similarly to quasi-resonant frequencies $\omega_{0}$.

Figure~\ref{fig:he-probs}a shows a clear transition of the ionization probability $P_{\rm ion}$ as a function of the chirp $\beta$ with an contrast of about 80\,\%.
Closely related, and almost complementary to this behavior, the excitation probability $P_{\rm ex}$ changes.
For chirp values $|\beta|>2$ these two quantities add up approximately to 1.
For smaller $\beta$ the initial 1s state does not get fully depleted, cf.\ black line in Fig.\,\ref{fig:he-probs}a. The symmetry $P_{\rm ini}({-}\beta)=P_{\rm ini}({+}\beta)$ follows
from a more general relation for pulses ``flipped'' in time 
\cite{suppl}.
Figure~\ref{fig:he-probs}b shows the ionization (and for completeness the excitation) probability as a function of laser intensity $I$ for two chirp parameters $\beta=\pm3$ and fixed pulse duration of $T=3$\,fs \cite{pulse}.
The graph covers the entire range from negligible ionization to full saturation.
Note, that this transition occurs at rather different intensities for positive and negative chirps $\beta$. 

Finally, Fig.\,\ref{fig:he-prob-spec} confirms the important role of two-photon absorption for
the chirp-sensitive ionization according to Eq.\,\eqref{eq:ampJ}: Long pulses are too weak for efficient two-photon processes such that the condition for perfect interference is missed by an increasing margin lowering the contrast of negative and positive chirped ionization. Short pulses weaken the prevalence of the resonant energy $k$ due to their large energy spread and the fact that also higher-order multi-photon processes can be realized due to the large intensity of the light. This is indicated by the larger deviation between full ionization and ionization by two photons (dashed lines) for short pulses. Moreover, ultrashort pulses activate the qualitatively different regime of non-adiabatic photo-ionization \cite{nisa+18}. 

To summarize: If rapid adiabatic passage is extended to include transitions to the continuum, the direction of the chirp decides if ionization is suppressed in favor of excitation with high contrast or vice versa. The phenomenon is universal for suitable
combinations of parameters which occur naturally, e.g., for intense VUV pulses exciting resonantly
the 1s-2p transition in helium. It is remarkable that the locking of the bound states due to the chirp
persists during the main part of the pulse almost unaffected by the laser-induced interaction with free electrons which act as an ordinary photon-dressed continuum. This may in general simplify in the future the approach to electron dynamics controlled by shaped light pulses.

One of us (US) gratefully acknowledges a lively discussion with A.\,Emmanouilidou, R.\,Moshammer, T.\,Pfei\-fer, and A.\,Saenz during the ``Atomic Physics Workshop 2017'' in Dresden. 
This work has been supported by the Deutsche Forschungsgemeinschaft (DFG) through the priority program 1840 ``Quantum Dynamics in Tailored Intense Fields''.

\def\articletitle#1{\emph{#1}.}

\end{document}